\documentclass[preprint,authoryear,12pt]{elsarticle}

\usepackage{graphicx}
\usepackage{epstopdf}
\usepackage{booktabs}
\usepackage{natbib} 
\usepackage{multirow}

\usepackage{amssymb}
 \usepackage{amsthm}

\usepackage{color,soul} 
\definecolor{lightred}{rgb}{1,.6,.6}
\sethlcolor{lightred}
\setulcolor{red}
\setstcolor{red}

\begin{document}

\begin{frontmatter}



\title{Co-movement of energy commodities revisited: Evidence from wavelet coherence analysis}


\author[utia,ies]{Lukas Vacha} \ead{vachal@utia.cas.cz}
\author[ies,utia]{Jozef Barunik\corref{cor2}} \ead{barunik@utia.cas.cz}

\cortext[cor2]{Corresponding author}

\address[utia]{Institute of Information Theory and Automation, Academy of Sciences of the Czech Republic, Pod Vodarenskou Vezi 4, 182 00, Prague, Czech Republic} 
\address[ies]{Institute of Economic Studies, Charles University, Opletalova 21, 110 00, Prague,  Czech Republic}

\begin{abstract}

In this paper, we contribute to the literature on energy market co-movement by studying its dynamics in the time-frequency domain. The novelty of our approach lies in the application of wavelet tools to commodity market data. A major part of economic time series analysis is done in the time or frequency domain separately. Wavelet analysis combines these two fundamental approaches allowing study of the time series in the time-frequency domain. Using this framework, we propose a new, model-free way of estimating time-varying correlations. In the empirical analysis, we connect our approach to the dynamic conditional correlation approach of \cite{engle2002} on the main components of the energy sector. Namely, we use crude oil, gasoline, heating oil, and natural gas on a nearest-future basis over a period of approximately 16 and $1/2$ years beginning on November 1, 1993 and ending on July 21, 2010. Using wavelet coherence, we uncover interesting dynamics of correlations between energy commodities in the time-frequency space.
\end{abstract}


\end{frontmatter}

\newpage
\section{Introduction}

Energy commodities play an important role in economic research, as they affect a wide range of markets and a variety of market participants operating on these markets. Study of the dynamics and statistical properties of energy commodities has become an important part of financial analysis since commodities became an additional tool for international diversification between stocks, bonds and currencies. Energy-related costs play a crucial role in the decision making of industrial companies and entrepreneurs. A proper understanding of the price dynamics and interconnections between commodities has become a fundamental concern. Moreover, energy commodity prices tend to have different and more extreme statistical properties than prices of other financial assets such as stocks, bonds, exchange rates, and corresponding derivatives. Prices of energy commodities are also influenced by various types of investors. Apart from standard financial investors, demand for energy commodities consists of industrial entities and power stations using the commodities to produce electricity.

The commodity markets are complex systems of interacting agents with different term objectives. Hence, time series resulting from this process are formed by a combination of different components operating at different frequencies. Standard time series econometric methods usually consider the frequency and time components separately. In this paper we introduce the wavelet approach, which allows us to study the frequency components of time series without losing the time information. The introduction of wavelets helps us uncover interactions which are hard to see using any other modern econometric method and which would otherwise stay hidden. Moreover, the wavelet analysis approach is model-free. This property makes it a very powerful tool in comparison with other methods that rely on parameters as well as the estimation method. We are not the first to apply wavelets in the analysis of commodities. \cite{davidson} proposed a form of semi-nonparametric regression based on wavelet analysis to study commodity price behavior. \cite{yousefi} used wavelets for prediction of oil prices. \cite{ConnorRossiter} were the first to estimate price correlations by basic scale decomposition of time series using wavelets on commodity markets. They use the discrete type of wavelet transformation, while in our analysis we follow the continuous transform approach. Contrary to \cite{ConnorRossiter}, we study the dynamics of correlations in the time-frequency space; our results are thus even more general. Recently, \cite{Naccache2011} used wavelets for the correlation analysis of oil prices and economic activity. However, that study was restricted to monthly data, with its main focus on longer cycles.

In recent years, applied research on the dynamics of energy commodities has grown significantly. Literature on the topic has been presented in three main categories -- research on the dynamics of specific commodity time series, research on the macroeconomic relationships of energy commodities, and research on co-movements between commodities. Several studies have examined the topic of co-movement of commodities. For example, \cite{Lanza2006} examined the dynamic conditional correlations in daily returns of WTI oil futures prices and found these to vary significantly. \cite{Grine2010} used cross-correlations to successfully construct a jump-diffusion model for pricing of cross-commodity derivatives of energy assets. Finally, \cite{Ghoshray2010} concluded that trends in energy prices change frequently and are hard to predict.

In our research, we contribute to the discussion of co-movements and use a novel approach which is much easier to interpret. The application of wavelets enables us to study the interdependence of energy time series in the time as well as frequency domains, providing a deeper understanding of possible dependencies. The main question of our analysis is whether the interconnection between studied commodity markets changes significantly in time and varies across different investment horizons. 

More specifically, we focus on dynamic correlations based on wavelet coherence between crude oil, gasoline, heating oil, and natural gas. By doing so, we analyze the evolution of these correlations in time as well as for different frequencies. This approach distinguishes between different types of investors with different investment horizons. Moreover, cyclical components, which are an important part of energy commodity dynamics, are also easily distinguishable and separated between various time periods. In addition to wavelet coherence in the time-frequency space, we propose a new way of estimating model-free time-varying correlations. To relate our results to the standard econometric literature, we connect our approach with the standard econometric approach of \cite{engle2002} -- dynamic conditional correlations from a multivariate GARCH model.

This paper is structured as follows. After a brief introduction to wavelets, Section 2 presents the wavelet coherence used for the estimation of local correlation in the time-frequency domain, as well as the phase differences used to characterize the relationship between the two time series. The comparison model -- DCC GARCH -- is also briefly introduced in this part. Section 3 introduces the data and provides results of the empirical analysis. Section 4  concludes by stating our main result, namely, a very interesting evolution of correlations between energy commodities. 

\section{Methodology}

\subsection{Wavelet analysis}

The wavelet transform offers localized frequency decomposition, providing information about frequency components. As a result, wavelets have significant advantages over basic Fourier analysis when the object under study is locally stationary and inhomogeneous -- see \cite{Gencay2002, PercivalWalden2000,Ramsay2002}. In our work we use continuous wavelet analysis tools, mainly wavelet coherence, measuring the degree of local correlation between two time series in the time-frequency domain, and wavelet coherence phase differences. Before introducing wavelet coherence, let us provide basic definitions of wavelet and wavelet transform.

\subsubsection{Wavelet}

A wavelet is a real-valued square integrable function, $\psi\in L^2(\mathbb{R})$\footnote{A function $x(t)$ is called a square integrable if $\int_{-\infty }^{\infty } x(t)^2 dt<\infty$.}, defined as:
\begin{equation}
\psi_{u,s}(t)=\frac{1}{\sqrt s}\psi \left( \frac{t-u}{s}\right),
\end{equation}
where the term $1/\sqrt{s}$ denotes a normalization factor ensuring unit variance of the wavelet, $\| \psi_{u,s} \|^2=1$. A wavelet has two control parameters, $u$ and $s$. The location parameter $u$ determines the exact position of the wavelet and the scale parameter $s$ defines how the wavelet is stretched or dilated. Scale has an inverse relation to frequency; thus lower (higher) scale means a more (less) compressed wavelet, which is able to detect higher (lower) frequencies of a time series. In addition, there are several conditions that a wavelet needs to satisfy. The most important is the admissibility condition, ensuring reconstruction of a time series from its wavelet transform. The admissibility condition is defined as
\begin{equation}
C_{\psi}=\int_{0}^{\infty }\frac{\mid\Psi(f)\mid^2}{f}df<\infty, 
\end{equation}
where $\Psi(f)$ is the Fourier transform of a wavelet $\psi(.)$. The admissibility condition implies that the wavelet does not have a zero frequency component and so the wavelet has zero mean, $\int_{-\infty }^{\infty }\psi (t)dt=0$. Further, the wavelet is usually normalized to have unit energy, i.e., $\int_{-\infty }^{\infty }\psi^2 (t)dt=1$, implying that the wavelet makes some excursion away from zero.

There is a large number of different wavelets. Each wavelet has its specific characteristics and is used for different purposes -- see \cite{PercivalWalden2000, Adisson2002}. In our analysis we use the Morlet wavelet, defined as:
\begin{equation}
\psi^M(t)=\frac{1}{\pi^{1/4}}e^{i\omega_0 t}e^{-t^2/2}.
\end{equation}
Parameter $\omega_0$ denotes the central frequency of the wavelet.  We set $\omega_0=6$, which is often used in economic applications -- see for example \cite{Conraria2008, RuaNunes2009}. The Morlet wavelet belongs to the family of complex or analytic wavelets, hence this wavelet has both real and imaginary parts, allowing us to study both amplitude and phase. 

\subsubsection{The continuous wavelet transform}

The continuous wavelet transform $W_x(u,s)$ is obtained by projecting a specific wavelet $\psi(.)$ onto the examined time series $x(t)\in L^2(\mathbb{R})$, i.e.,
\begin{equation}
W_x (u,s)=\int_{-\infty}^\infty x(t)\frac{1}{\sqrt s}\overline{\psi \left( \frac{t-u}{s}\right)} dt.
\label{eq7}
\end{equation}
An important feature of the continuous wavelet transform is the ability to decompose and then subsequently perfectly reconstruct a time series $x(t)\in L^2(\mathbb{R})$:
\begin{equation}
x(t)=\frac{1}{C_\psi}\int_0^\infty \left[\int_{-\infty}^\infty  W_x (u,s) \psi_{u,s}(t) du \right] \frac{ds}{s^2} ,\hspace{5 mm}s>0.
\end{equation}
Furthermore, the continuous wavelet transform preserves the energy of the examined time series,  
\begin{equation}
\| x \|^2 =\frac{1}{C_\psi}\int_0^\infty  \left[\int_{-\infty}^\infty  \left| W_x (u,s)\right|^2 du \right] \frac{ds}{s^2}.
\end{equation}
We use this property for the definition of wavelet coherence, which measures the size of the local correlation between two time series.

\subsubsection{Wavelet coherence}

To be able to study the interaction between two time series, we need to introduce a bivariate framework called wavelet coherence. For the proper definition of the wavelet coherence, we need to introduce the cross wavelet transform and cross wavelet power first. \cite{TorenceCompo98} defined the cross wavelet transform of two time series $x(t)$ and $y(t)$ as
\begin{equation}
W_{xy} (u,s) = W_x (u,s) W_y^* (u,s),
\end{equation}
where $W_x (u,s)$ and $W_y (u,s)$ are continuous wavelet transforms of $x(t)$ and $y(t)$, respectively, $u$ is a position index, and $s$ denotes the scale, while the symbol $^*$ denotes a complex conjugate. The cross wavelet power can easily be computed using the cross wavelet transform as $|W_{xy} (u,s)|$. The cross wavelet power reveals areas in the time-frequency space where the time series show a high common power, i.e., it represents the local covariance between the time series at each scale.

The wavelet coherence can detect regions  in the time-frequency space where the examined time series  co-move, but do not necessarily have a high common power. Following the approach of \cite{TorrenceWebster99}, we define the squared wavelet coherence coefficient as: 
\begin{equation}
R^2 (u,s)=\frac{|S(s^{-1}W_{xy} (u,s))|^2}{S(s^{-1}|W_x (u,s)|^2) S(s^{-1}|W_y (u,s)|^2)},
\end{equation}
where $S$ is a smoothing operator\footnote{Without smoothing, the wavelet coherence equals one at all scales. Smoothing is achieved by convolution in both time and scale. The time convolution is performed with a Gaussian window, while the scale convolution is done with a rectangular window -- see \cite{Grinsted2004}.}. The squared wavelet coherence coefficient is in the range $0\le R^2 (u,s) \le1$. Values close to zero indicate weak correlation, while values close to one provide evidence of strong correlation. Hence, the squared wavelet coherence measures the local linear correlation between two stationary time series at each scale and is analogous to the squared correlation coefficient in linear regression.

Since the theoretical distribution for the wavelet coherence is not known, we test the statistical significance using Monte Carlo methods. In the testing procedure, we follow the approach of \cite{Grinsted2004} and \cite{TorenceCompo98}.

The use of wavelets brings with it the difficulty of dealing with boundary conditions on a dataset with finite length. This is a common problem with any transformation relying on filters. In our paper, we deal with this problem by padding the time series with a sufficient number of zeroes. The area where the errors caused by discontinuities in the wavelet transform cannot be ignored, i.e., where edge effects become important, is called the cone of influence\footnote{The cone of influence is highly dependent on the type of wavelet used -- see \cite{TorenceCompo98}.} \citep{Grinsted2004}. The cone of influence lies under a cone which is bordered by a thin black line.

\subsubsection{Phase}
To complete our analysis, we also use wavelet coherence phase differences showing us details about the delays in the oscillation (cycles) between the two time series under study. Following \cite{TorrenceWebster99} we define the wavelet coherence phase difference as:
\begin{equation}
\phi_{xy} (u,s)=\tan ^{-1}\left( \frac{\Im \{S(s^{-1}W_{xy} (u,s))\}}{\Re \{S(s^{-1}W_{xy} (u,s))\} }\right).
\end{equation}
Phase is indicated by arrows on the wavelet coherence plots. A zero phase difference means that the examined time series move together. The arrows point to the right (left) when the time series are in-phase (anti-phase) or are positively (negatively) correlated. Arrows pointing up means that the first time series leads the second one by $90^{\circ}$, whereas arrows pointing down indicates that the second time series leads the first one by $90^{\circ}$. Usually we have a mixture of positions, for example, an arrow pointing up and right means that the time series are in phase, with the first times series leading the second one. 

\subsection{DCC GARCH analysis}

We would like to provide a connection of this non-traditional analysis to standard econometric tools. In fact, we step back from the time-frequency domain into the time domain to provide this connection. For this purpose, we compute local time-varying correlations from the wavelet coherence and compare them with the well-established multivariate concept of Dynamic Conditional Correlation Generalized Autoregressive Conditional Heteroscedasticity (DCC GARCH) of \cite{engle2002}. In this part, we will provide a very basic overview of the model.

The DCC estimator is a logical extension of the constant conditional correlation (CCC) model introduced by \cite{bollerslev90}. In Bollerslev's model, correlation matrix $R$ is constant: $H_t=D_t R D_t,$ where $D_t=diag\{\sqrt{h_{i,t}}\}$ and $h_{i,t}$ represents the $i$-th univariate (G)ARCH$(p,q)$ process, and $i=1,\dots,n$ at time $t=1,\dots,T$. \cite{engle2002} allowed $R$ to vary in time $t$, thus
\begin{equation}
H_t=D_t R_t D_t.
\end{equation}
The correlation matrix is then given by the transformation
\begin{equation}
R_t=diag(\sqrt{q_{11,t}},\dots,\sqrt{q_{nn,t}}) Q_t diag(\sqrt{q_{11,t}},\dots,\sqrt{q_{nn,t}}),
\end{equation}
where $Q_t=(q_{ij,t})$ is
\begin{equation}
Q_t=(1-\alpha-\beta) \overline{Q}+\alpha\eta_{t-1}\eta'_{t-1}+\beta Q_{t-1},
\label{eqQ}
\end{equation}
where $\eta_t=\epsilon_{i,t}/\sqrt{h_{i,t}}$ are the standardized residuals from the (G)ARCH model, $\overline{Q}=T^{-1}\sum{\eta_t \eta'_t}$ is a $n\times n$ unconditional variance matrix of $\eta_t$, and $\alpha$ and $\beta$ are non-negative scalars such that $\alpha+\beta<1$.

To estimate DCC GARCH, we use the standard procedure proposed by \cite{engle2002}.

\section{Empirical results}

\subsection{Data description}
To study the dependence between energy markets, we use the main components of the energy sector according to the Continuous Commodity Index constructed by the Commodity Research Bureau. Namely, we use crude oil, gasoline, heating oil, and natural gas on a nearest-future basis. The energy prices were collected on a daily basis over a period of approximately 16 and $1/2$ years beginning on November 1, 1993 and ending on July 21, 2010. Altogether, the sample includes 3573 daily prices for each commodity. The data were obtained from the Pinnacle Data Corporation. Figure \ref{fig:WTC} shows the normalized plots of the prices and Table \ref{stats} in Appendix A provides descriptive statistics for the logarithmic returns used in the analysis.

In Figure \ref{fig:WTC} we can directly observe that heating oil and crude oil are closely related. Gasoline departs from this relationship but still has some common trends during several specific periods. Natural gas seems to have the weakest relation to the other commodities. Let us explore these dependencies in more detail. We will first use the simple unconditional correlation coefficient of the returns for the whole data set, and then we will look at the evolution of the local correlations using wavelet coherence and connect the wavelet coherence results using the standard DCC GARCH approach.

\begin{figure}
\centering
\includegraphics[scale=0.5]{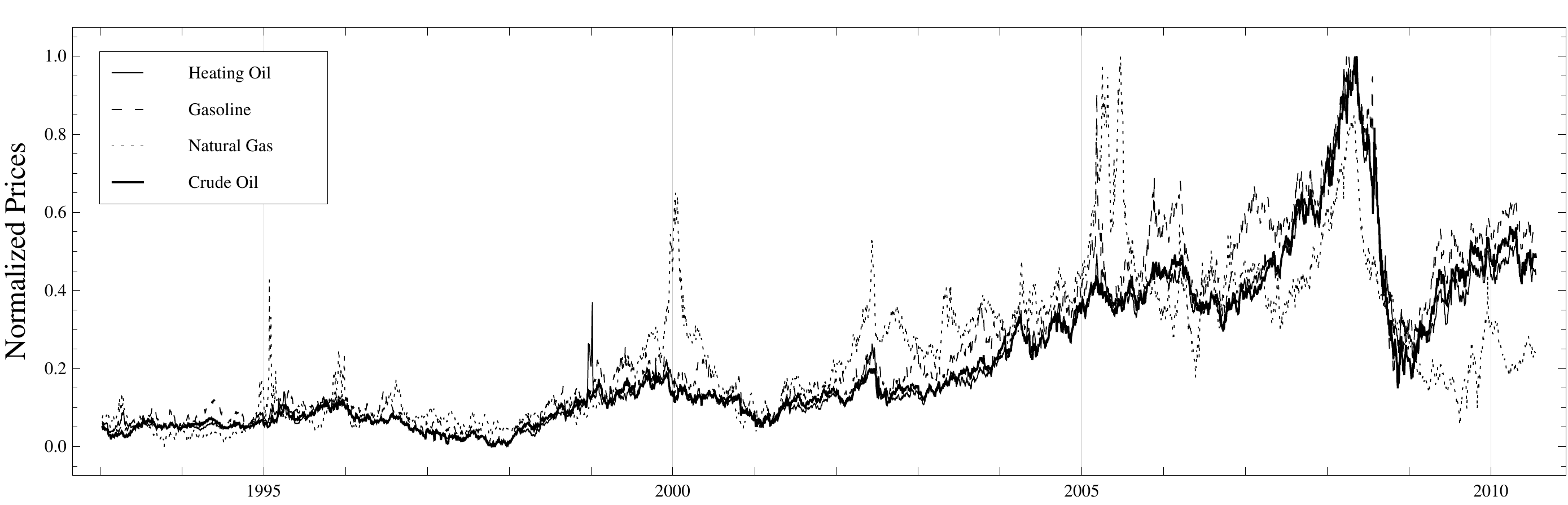}
\caption{Normalized prices of heating oil, gasoline, natural gas, and crude oil.}
\label{fig:WTC}
\end{figure}

Unconditional correlations in Table \ref{correlations} show that heating oil, crude oil, and gasoline are closely related, providing high correlations. Heating oil has the strongest relationship with crude oil. Gasoline is related to crude oil more strongly than to heating oil, but all three pairs show positive correlations of around 0.6. The dependence of natural gas on the rest of the group is the weakest, from the point of view of the correlations.

\subsection{Evidence from the wavelet coherence}

Unconditional correlations provide evidence of strong dependencies in the energy commodities. However, the time span of our study is quite long, so it may be interesting to see how the correlations develop in time. It may also be of interest to learn whether the dependencies vary across different frequencies, i.e., if there are stronger dependencies in the longer or shorter investment horizons. The wavelet coherence approach will be used as a tool allowing us to study the dependence in time as well as frequency domains.

\begin{figure}
\centering
\includegraphics[scale=0.48]{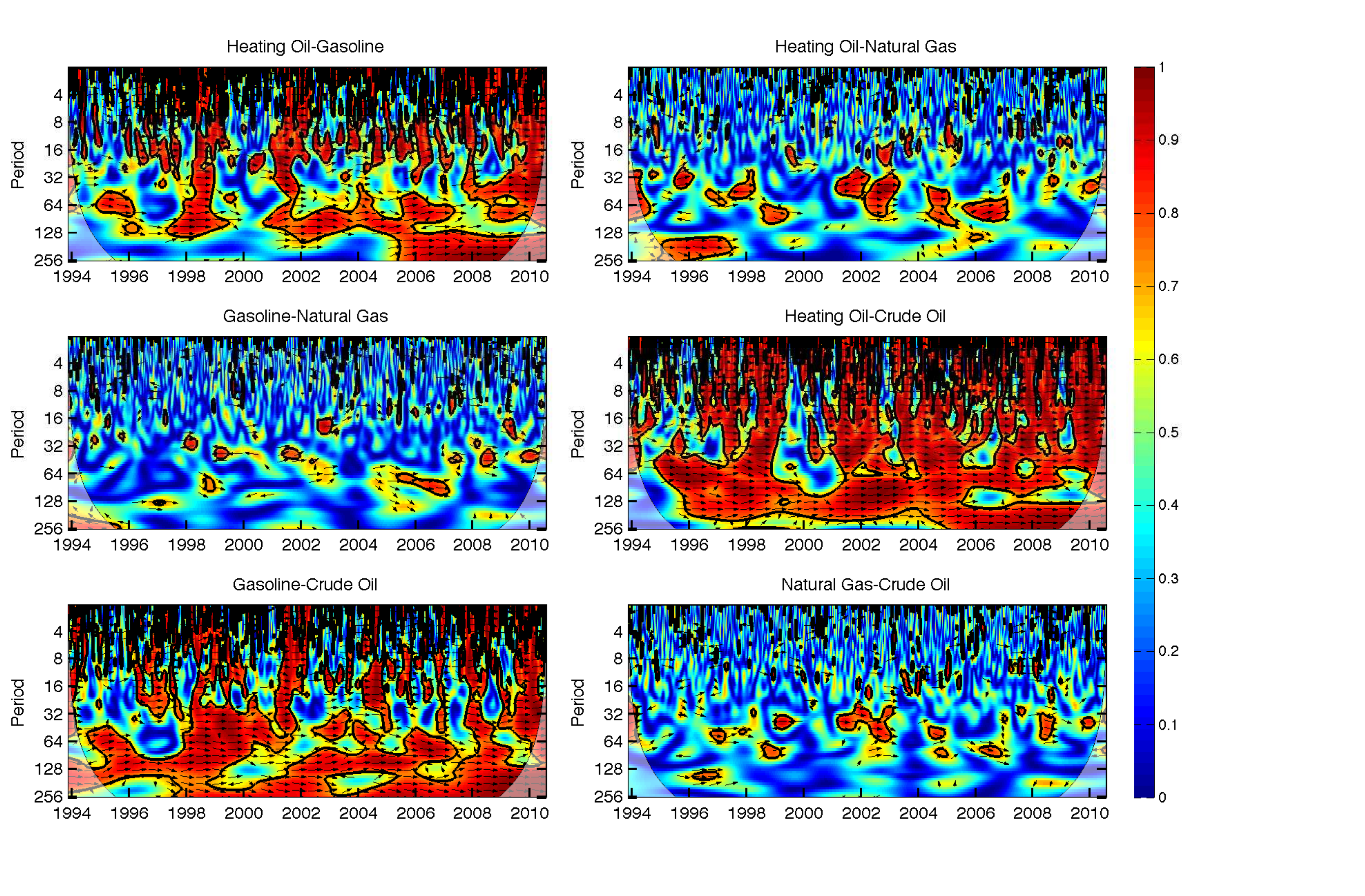}
\caption{Wavelet coherence of heating oil, gasoline, natural gas and crude oil pairs. The horizontal axis shows time, while the vertical axis shows the period in days. The warmer the color of a region, the higher the degree of dependence between the pair.}
\label{fig:WTC2}
\end{figure}

To assess the statistical significance of the local correlations in the time-frequency space, we use Monte Carlo simulations. Figure \ref{fig:WTC2} shows the estimated wavelet coherence and the phase difference for all examined pairs of indices from scale 1 (one day) up to a scale of 256 (approximately one market year). Time is shown on the horizontal axis, while the vertical axis refers to frequency; the lower the frequency, the higher the scale. The wavelet coherence finds the regions in time-frequency space where the two time series co-vary. Regions inside the black lines plotted in warmer colors represent regions with significant dependence. The colder the color is, the less dependent the series are. Cold regions outside the significant areas represent time and frequencies with no dependence in the commodities. Thus, we can clearly see both the frequency and the time intervals where the commodities move together significantly. A continuous wavelet transform at any given point uses the information of neighboring data points, so areas at the beginning and end of the time interval should be interpreted with caution, as discussed in previous sections. This is also the reason we include only scales up to 256.

From analysis of the wavelet coherence, we can observe very interesting results. A first glance confirms our findings from the previous analysis. The three commodities of heating oil, gasoline, and crude oil show the strongest dependence. Heating oil is strongly related to crude oil over many periods and at many frequencies. Gasoline is also quite strongly related to crude oil. The dependence of heating oil and gasoline turns out to be restricted to only some periods. For example, around 1998, and 2001, these two were very closely related at almost all frequencies, but around the year 2000, there seems to be no dependence at all. This interesting finding may have its origins in the period of recessions with falling prices. The result will be discussed in the following text in greater detail. 

We can thus see that the dependence is highly dynamic as it varies in time. Phases, represented by arrows, do not provide any additional value to the analysis,  as they point to the right most of the time, meaning that the significant local correlations are positive, but no commodity is leading (affecting the other one). There are short periods with changing phases and where one commodity seems to be leading the other, but these are not consistent so we cannot conclude that there is any directional influence. When we compare these three indices to natural gas, we can see very small significant areas of dependence.

\subsection{Connection to the standard analysis in time domain}

The wavelet coherence measures co-movement of two time series in the time-frequency space. Its main advantage is the ability to decompose the time-varying co-movement into different investment horizons. Since the wavelet coherence can be interpreted, in some sense, as a measure of local correlation, we would like to connect the obtained results to standard econometric analysis, usually performed in the time domain only. In this section, we connect the time-varying co-movement (estimated with the wavelet coherence) with the DCC GARCH model, which we borrow as an representative of standard methods\footnote{We perform only a simple-estimate comparison: this paper is not meant to provide a detailed discussion of the differences and capabilities of these models.}.

\begin{table}[h]
\footnotesize
\begin{center}
\begin{tabular}{@{} lllll @{}}
\toprule
 Ê& & Heating oil & Gasoline & Natural gas \\
\cmidrule{3-5}
			& & & & \\
 \multirow{3}{*}{Gasoline} & Unconditional & 0.568	& Ê& Ê\\
 			& DCC & 0.638	(0.137) Ê Ê& Ê & Ê\\
			& WTC & 0.541 (0.180) Ê Ê& Ê & Ê\\
			& & & & \\
 \multirow{3}{*}{Natural gas} & Unconditional & 0.147 & 0.083 & Ê\\
 			 Ê & DCC & 0.172 (0.0235) & 0.101 (9.188$\times10^{-12}$)	& Ê \\
 			 Ê& WTC & 0.268 (0.130) Ê&	0.181 (0.123)		& Ê \\
			& & & & \\
 \multirow{3}{*}{Crude oil} Ê& Unconditional & 0.643 & 0.612 & 0.077 Ê\\
			& DCC & 0.749 (0.141) & 0.645 (0.123) & 0.109 (0.025) \\
			& WTC & 0.708 (0.151) & 0.626 (0.131) & 0.186 (0.129) \\
\bottomrule
\end{tabular}
\caption{Correlation matrix - comparison of unconditional correlation coefficient with mean value of correlations obtained from the DCC GARCH and the wavelet coherence (WTC), with their standard deviations in the parentheses.}
\label{correlations}
\end{center}
\end{table}

To connect the two approaches, we have to reduce the wavelet coherence to the time dimension only. More precisely, we compute the  correlation coefficient based on the squared wavelet coherence coefficient for each time $t$ by simply averaging wavelet coherence (those not significant at a 95 \% level of significance are taken to be zero) at all scales for the particular time $t$ and taking the square root of this average. This reduces the wavelet coherence analysis to the time-varying correlations which are computed without the use of models. 

The squared wavelet coherence coefficients are in absolute value, hence negative correlations cannot be seen directly. While the wavelet coherence is performed with a complex wavelet, we obtain phase positions of the examined time series which might be used for distinguishing between positive and negative correlations. In our particular datasets, there are no negative correlations (anti-phase positions of time series); hence we do not need to compensate for the negative correlation.

Let us see how the time-varying correlations from the wavelet coherence are related to the correlations from the DCC GARCH. Table \ref{correlations} provides comparison of both methods to unconditional correlation. More precisely, Table \ref{correlations} compares the mean values of the correlation using different methods for the whole sample. For the three most correlated pairs, the estimates of correlation from the wavelet coherence are closer to the unconditional correlations than the estimates from the DCC GARCH. The opposite is true for the other three, least correlated, pairs. In the cases with very low unconditional correlation, DCC GARCH seems to converge to a constant correlation, which is close to an unconditional one. Still, except for the heating oil - crude oil pair, all unconditional correlations are within the standard deviation of arithmetic mean of correlations from the wavelet coherence as well as the DCC GARCH. These results point to the correctness of  both approaches\footnote{It should be noted that forecasting with the aid of wavelet coherences is difficult in the present setting due to the nature of wavelets.}.

\begin{figure}
\centering
\includegraphics[width=5.5in]{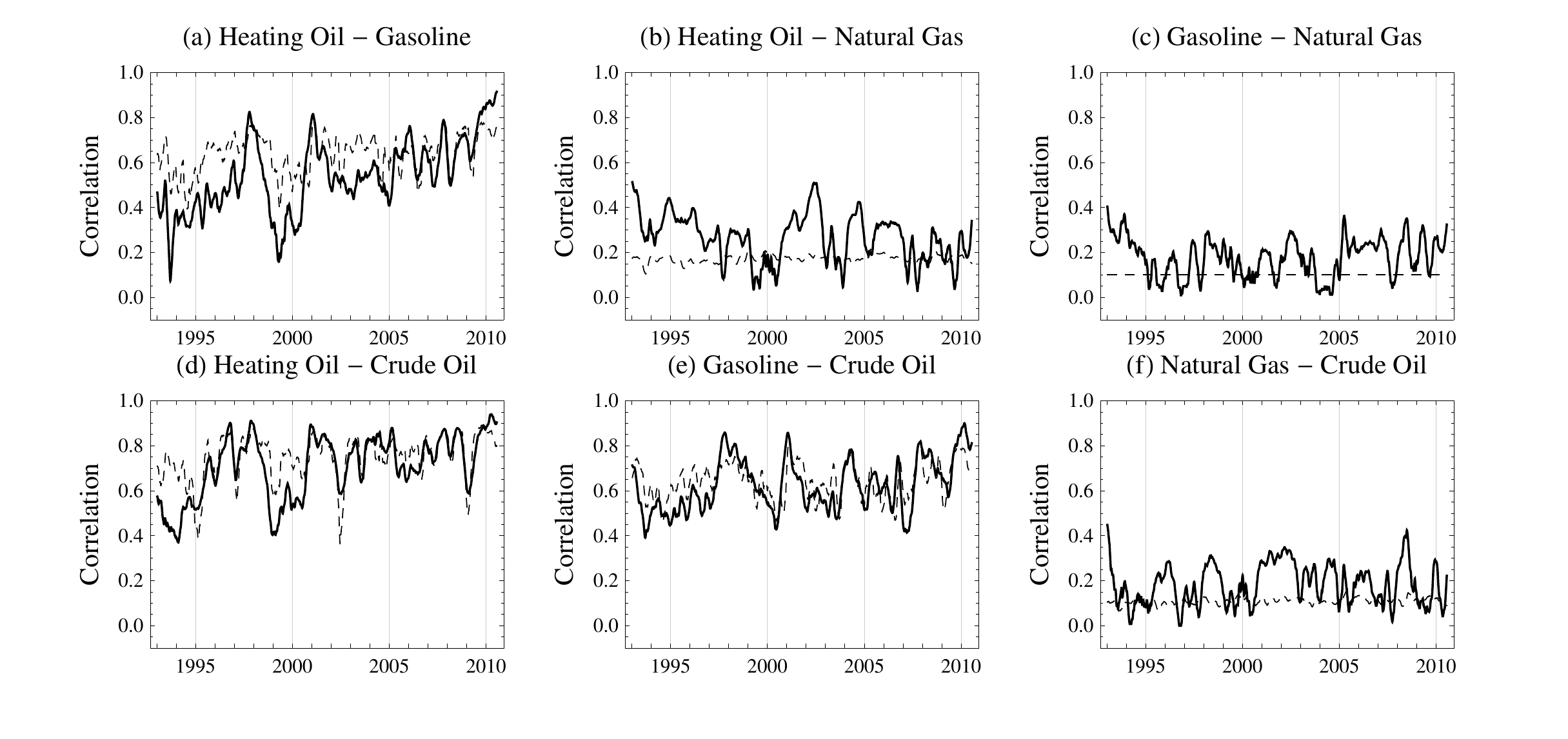}
\caption{Comparison of correlations between the DCC GARCH (dashed line) and time-varying correlations from the wavelet coherence (black line) for heating oil, gasoline, natural gas, and crude oil pairs. The results are smoothed using 50-day moving averages}
\label{fig:Comparison}
\end{figure}

Figure \ref{fig:Comparison} provides comparison of time-varying correlations using both approaches\footnote{We use 50-day moving averages for illustration.}. Since we use the comparison to DCC GARCH only as an illustration of how wavelet coherence is related to standard methods, we do not concentrate on the differences between approaches; rather, we discuss the findings in the following section.

\subsection{Time-frequency dynamics of the co-movement}

The time-varying dynamics of correlations is confirmed using both approaches. But the wavelet coherence provides a rather complex insight into this dynamics. Turning back to Figure \ref{fig:WTC2}, we can see that co-movement changes not only in time, but is different also at various investment horizons. The study of the pair of heating oil and gasoline is a good example:

In 1998, there was a large increase in the correlation of this pair, up to 0.8, confirmed by both DCC GARCH as well as wavelet coherence methods. Wavelet coherence plotted in Figure \ref{fig:WTC2} shows that this strong dependence was present for all investment horizons from the longer ones of several months, up to the shorter, daily ones. In the following two years, the correlation dropped quickly. The wavelet coherence plot again brings the very interesting insight that the only dependence present in this pair around the year 2000 was a short-term one -- up to one week. After the year 2000, the dependence again increased rapidly to the 0.8 level during the following year. The latest development of dependence among the three most correlated commodities -- heating oil, gasoline, and crude oil -- is very similar. Correlations between them increased rapidly to the 0.8 levels at the beginning of 2009. 

For these three most correlated commodities, the periods of high coherence around the years 1998 and 2001 are closely related to periods of recession with falling prices. Specifically, they relate to the Asian financial crisis in 1998-2000, the 9/11 terrorist attacks, and the resulting fear in the markets in 2001-2002. Finally, the current financial crisis of the period 2008-2010 has shown a similar behavior. All three commodities again have significant coherences through all investment horizons.

While we can see that the dynamics of correlation is changing rapidly in time, we can also see that the dynamics change in frequency as well. For example, the gasoline - crude oil pair shows a large dependence for investment horizons of several months during the years 1997-2001, while the short-term dependence seems to be very low during this period. The most correlated pair, heating oil - crude oil, shows the dynamics of the dependence in both time and frequency as well. The longer-term (up to six months) investors  believe that this dependence  is very high, while in the short-term investment horizons, the dependence varies over time much more quickly.

We can compare wavelet coherence with the findings of entropy analysis published recently by \cite{Martina2011}. The authors show that for crude oil, the periods with economic downturns exhibit a reduced market complexity in terms of reduced entropy levels. We observe similar behavior, because these periods are characterized by falling prices, i.e., movement in one direction. As a consequence, the complexity is reduced, and in the multivariate setting the co-movement is high and significant for a large number of investment horizons.

\section{Concluding remarks}

In this paper, we contribute to the literature on co-movement on the energy markets by researching the interconnections between the main components of the energy sector in the time-frequency space. The novelty of our approach lies in studying the co-movement of energy markets in the time-frequency space for the first time and comparing the results to the standard econometric tools for studying relationships between markets.

The main finding of this paper is that some energy pairs show strong dynamics in co-movement in time during various investment horizons. The results suggest that when looking at the dependence of energy markets, one should always keep in mind its time-varying nature and look at it for various investment horizons. While the strongest dependence occurs during the periods of sharp price drops, it seems that the periods of recession creating fear in the markets imply a much higher downside risk to a portfolio based on these commodities. This inefficiency of the energy market is muted after recovery from the recession. 

Wavelet coherence also uncovered long cycles (64 to 128 days) in heating oil - crude oil pair that were also present in the periods outside of recession, or more precisely, periods of stable growth. Dependence on higher frequency cycles changes considerably in time. Still, the three commodities, heating oil, gasoline and crude oil strongly co-move, thus for the manager willing to keep a well diversified portfolio, the trio will imply great exposure to risk. On the other hand, natural gas seems to be unrelated to all three commodities for all investment horizons as well as the studied time periods.

In conclusion, we uncover some interesting dynamics of the co-movement between energy markets in time as well as various investment horizons. Our findings are model-free and provide the possibility of new research on financial risk modeling, as they show that dynamic diversification is required in order to preserve a higher profit.

\section*{Acknowledgements}
We would like to thank Aslak Grinsted for providing us with the MATLAB wavelet coherence package. The authors would like to thank the anonymous referees whose suggestions helped to improve this paper substantially. The support from the Czech Science Foundation under Grants 402/09/0965, 402/09/H045, and 402/10/1610 and from the Department of Education MSMT 0021620841 is gratefully acknowledged.

\newpage
\section*{Appendix A}
\label{Appendix A}

\begin{table}[h]
\footnotesize
\begin{center}
\begin{tabular}{@{} lcccc @{}}
\toprule
 & Heating oil & Gasoline & Natural gas & Crude oil \\
\midrule
 Mean & 0.000368314 & 0.000400265 & 0.000211919 & 0.000412954 \\
 St.dev & 0.029718 & 0.0320496 & 0.049457 & 0.0277684 \\
 Skewness &  -1.28498 & -0.284786 & 0.178664 & -0.242562  \\
 Kurtosis & 34.5392 & 6.79143 & 22.049 & 8.01979 \\
 Min & -0.520032 & -0.222545 & -0.5187 & -0.200321 \\
 Max & 0.244974 & 0.230015 & 0.659246 & 0.181293 \\
 \bottomrule
\end{tabular}
\caption{Descriptive statistics of the daily logarithmic returns for heating oil, gasoline, natural gas, and crude oil in the period of November 1, 1993 to July 21, 2010.}
\label{stats}
\end{center}
\end{table}

\begin{table}[h]
\footnotesize
\begin{center}
\begin{tabular}{@{} lcc @{}}
\toprule
 & Coefficient & Standard Deviation \\
\midrule
 $\alpha_{DCC}$ & 0.04435448 & 0.01306025  \\
  $\beta_{DCC}$ & 0.90056051 & 0.03265073  \\
  \midrule
 Log.Likelihood & 82605.52 &\\
 \bottomrule
\end{tabular}
\caption{Coefficient estimates of multivariate DCC GARCH for heating oil, gasoline, natural gas, and crude oil in the period of November 1, 1993 to July 21, 2010. $\alpha_{DCC}$ and $\beta_{DCC}$ are coefficients from Equation \ref{eqQ}}
\label{DCCresults}
\end{center}
\end{table}

\newpage
\bibliography{Energy}
\bibliographystyle{chicago}

\end{document}